\documentclass[preprint]{aastex}
\usepackage{epsfig}

\shortauthors{Abbett, Fisher \& Fan} 
\shorttitle{Effects of Rotation on Rising Omega-loops}

\begin{document}

\title{The Effects of Rotation on the Evolution of Rising 
Omega-loops in a Stratified Model Convection Zone \\}
\author{W. P. Abbett and G. H. Fisher}
\affil{Space Sciences Laboratory, University of California,
    Berkeley, CA 94720-7450}
\and
\author{Y. Fan\altaffilmark{1}}
\affil{HAO, National Center for Atmospheric Research, P.O. Box
   3000, Boulder, CO 80307}
\email{yfan@hao.ucar.edu}
\altaffiltext{1}{The National Center for Atmospheric Research is
sponsored by the National Science Foundation.}

\begin{abstract}
We present three-dimensional MHD simulations of buoyant 
magnetic flux tubes that rise through a stratified model 
convection zone in the presence of solar rotation.  The
equations of MHD are solved in the anelastic approximation, 
and the results are used to determine the effects of solar 
rotation on the dynamic evolution an $\Omega$-loop. 
We find that the Coriolis force significantly suppresses
the degree of fragmentation at the apex of the loop during
its ascent toward the photosphere.  If the initial axial 
field strength of the tube is reduced, then, in the absence 
of forces due to convective motions, the degree 
of apex fragmentation is also reduced.  We show that 
the Coriolis force slows the rise of the tube, and induces 
a retrograde flow in both the magnetized and unmagnetized 
plasma of an emerging active region.  Observationally, we
predict that this flow will appear to originate at the leading 
polarity, and will terminate at the trailing polarity. 

\end{abstract}


\section{Introduction}
Active regions represent the largest observable concentrations of 
magnetic flux on the Sun.  The magnetic field within active
regions exhibits a bipolar structure, which suggests that
they are the tops of large $\Omega$-shaped loops which have risen
through the convection zone and have emerged through the photosphere.
On average, active region bipoles are oriented nearly parallel 
to the East-West direction (Hale's Polarity Law).  This suggests 
that the subsurface field geometry is approximately toroidal. 
Since active regions appear to be the tops of large, rising
$\Omega$-loops, and since Hale's law persists for years in
a given solar cycle, we can infer that the toroidal flux
must exist well below the solar surface, and that the magnetic
flux must reside in a region where it is relatively free from disruption
by buoyant instabilities or convective motions 
\citep{svb82,vb82,fms93,fms95}.  However, if the magnetic flux
is embedded in the radiative zone (well below the convective zone) 
it would be so stable that it could not emerge through the surface
on the time scale of a solar cycle \citep{p79,vb82}.  This
implies that the toroidal field likely resides in the ``convective 
overshoot region'' --- a thin, slightly convectively
stable layer located at the interface of the radiative and convective
zones.  This layer also appears to coincide with the ``tachocline''
\citep{k96,c99}, where the solid body rotational behavior of the 
radiative interior transitions into the differential rotation behavior
observed in the convection zone.  Thus, the toroidal field seems
to be generated and stored in the overshoot layer.
\citep{gmd89,dg91,p93,mc97,d97,dc99}.

We now have a basic picture of how active regions might form: 
A portion of the toroidal flux layer 
eventually succumbs to one of several possible instabilities 
(eg. \citealt{cms95,ff96,f00,wpmh00}), 
and forms a magnetically buoyant tube that begins to rise through the 
convection zone toward the photosphere. The top of the emerging
flux loop is then observed at the photosphere as a bipolar magnetic 
region.  The simplest way to model this picture is with
the ``thin flux tube'' approximation (see \citealt{d76,rw78,s81}).
In this simple model, it is assumed that magnetic flux tubes move 
through a field-free plasma, that the cross-section of the tube is much
smaller than all other relevant length scales of the problem, and 
that pressure balance is always maintained across the tube.  
Given these assumptions, one can derive an equation of motion 
for a one-dimensional tube moving through a three dimensional
medium. This model has had great success in explaining
many of the observed properties of active regions, including
(but not limited to) the variation and scatter of active region 
tilt as a function of solar latitude 
\citep{dh93,dc93,ffd93,cms95,ffh95,lf96,ff96}, 
the asymmetries in the orientation of active regions just 
after emergence \citep{vp90,mcs94}, and the latitudinal
variation and fluctuations of field line twist from active region
to active region \citep{lfp98}.

However, from the start, the thin flux tube approximation
assumes that the magnetic field will retain its ``tube-like''
identity; that is, it will not fragment or disperse as it
evolves.  This assumption is called into question by recent
two-dimensional MHD simulations which show 
that unless flux tubes possess a critical amount of initial 
field line twist along the tube --- and observations of emerging 
active regions suggest that many do not \citep{pcm95,lfp98,llp99}
--- the tubes will fragment (break apart) before they are able to 
emerge through the surface \citep{llf96,mie96,em98,fzl98}.  
Recently, \citet{aff00} used three-dimensional MHD simulations 
to demonstrate that the degree of cohesion at the apex of an 
emerging $\Omega$-loop depends strongly on the 
three-dimensional geometry of the loop.  If the curvature of 
the $\Omega$-loop at its apex is relatively large (ie. if the 
loop is tall and narrow), then the apex will emerge through 
the photosphere more cohesively than if the curvature is small 
(ie. if the $\Omega$-loop is short and wide).  \citet{dn98} and 
\citet{dan99} also performed detailed three-dimensional MHD simulations 
of buoyant magnetic flux tubes.  All of these 3-D studies 
still require a significant amount of initial field line twist 
in order to suppress the eventual fragmentation of the tube near 
the loop apex.  Is there a way out this quandary which
\citet{aff00} refer to as ``Longcope's paradox''?

In this paper, we investigate how solar rotation affects the 
dynamic evolution of buoyant magnetic flux tubes.  We find
that the Coriolis force plays a dynamically important role during 
the tube's ascent, and acts to maintain the tube's cohesion
even in the absence of initial field line twist, possibly
resolving Longcope's paradox.  In Section~\ref{method}
we briefly describe our computational methodology, discuss
how we include the Coriolis force in our computational model,
and define the relevant physical parameters of our problem.
In Section~\ref{results} we analyze the results of our
simulations, and develop a very simple model that effectively
characterizes the simulation results pertaining to 
flux tube cohesion.  In this section we also comment on possible 
observational consequences of the simulations.  
Finally, we present our summary and conclusions in 
Section~\ref{conclusions}.

\clearpage

\section{Method}\label{method}
We use the methods described in \citet{aff00} and \citet{fzl99}
to solve the three-dimensional equations of MHD in the anelastic
approximation \citep{op62,g69,lf99}.  The anelastic equations
result from an expansion of the equations of compressible MHD
about a zeroth order reference state (which we take 
to be a field-free, adiabatically stratified polytrope of index 
$m=1.5$). In this approximation, the higher order time derivative 
of the perturbed density is neglected, effectively filtering out 
fast-moving acoustic waves.  This is computationally advantageous, as 
it allows for much larger timesteps than would be possible in a 
fully compressible calculation.  This approximation is valid in 
sub-surface regions where the acoustic Mach number is small 
($M \equiv v/c_\mathrm{s} \ll 1$), and where the local sound speed 
($c_\mathrm{s}$) greatly exceeds the Alfv\'en speed ($v_{A}$) 
of the plasma \citep{g84}.  A detailed theoretical discussion of the 
anelastic formulation can be found in \citealt{lf99} (and references
therein).

The anelastic equations in a non-rotating reference frame are given 
by equations [1]-[7] in \citet{aff00} (hereafter AFF).  To consider the
effects of solar rotation, we adjust the anelastic momentum equation
to include non-inertial terms (we note, here, that due to a typesetting 
error in the final version of AFF, a cross-product symbol was omitted 
in the Lorentz force term of equation [2]). To consider the
effects of solar rotation, we adjust the anelastic momentum equation 
to include non-inertial terms.  In this study, we neglect centrifugal 
effects and concentrate only on the effects of the Coriolis force
due to solid-body rotation.  In this approximation, the 
anelastic momentum equation (in the rotating reference frame) 
becomes: 
\begin{equation}\label{momentum}
\rho_0\left(\frac{\partial\mathbf{v}}{\partial{t}}+
   \mathbf{v}\cdot\mbox{\boldmath$\nabla$\unboldmath}
   \mathbf{v}\right)=-\mbox{\boldmath$\nabla$\unboldmath}
   p_1+\rho_1\mathbf{g}-2\rho_0\left(
   \mbox{\boldmath$\Omega$\unboldmath}\times\mathbf{v}\right)+
   \frac{1}{4\pi}\left(\mbox{\boldmath$\nabla$\unboldmath}
   \times\mathbf{B}\right)\times\mathbf{B}+
   \mbox{\boldmath$\nabla$\unboldmath}\cdot\mathbf{\Pi} \, .
\end{equation}
Here, $\rho_0$ refers to the density stratification of the
zeroth order reference state, while $\rho_1$, $p_1$, $\mathbf{v}$, 
and $\mathbf{B}$ represent the density, pressure, velocity and 
magnetic field perturbations.  The viscous stress tensor is given by
$\Pi_{ij}\equiv\mu (\partial v_i/\partial x_j + \partial v_j/
\partial x_i - 2/3(\mbox{\boldmath$\nabla$\unboldmath}
\cdot\mathbf{v})\delta_{ij})$, where
$\mu$ is the coefficient of viscosity and $\delta_{ij}$ is the 
Kronecker delta function. \boldmath $\, \Omega$ \unboldmath denotes 
the angular velocity about the Sun's axis 
of rotation.  Both the acceleration due to gravity, 
\boldmath ${\rm g} \mbox{\unboldmath$\,= -$g}\,\hat{z}$\unboldmath, 
and the coefficient of viscosity are assumed constant in our 
calculations. 

We solve the non-dimensional form of the anelastic equations in
the modified, local \emph{f}-plane approximation of \citet{bht96}.  
This approach is similar to the standard \emph{f}-plane approximation
of geophysical fluid dynamics \citep{p87}, except that the transverse
components of the Coriolis force within the rectangular domain are
retained.  Since $\Omega$ is assumed constant throughout the
Cartesian box (positioned at a given latitude), we can apply 
simple, periodic boundary conditions in the horizontal directions, 
and stress-free, non-penetrating conditions at the upper and lower 
boundaries.  Further simplification is possible by noting that the 
momentum density (in the anelastic approximation) 
and magnetic field vectors are both divergence-free.  
This allows us to define scalar potentials such that 
$\mathbf{B}\equiv\mbox{\boldmath$\nabla$\unboldmath}
\times\mbox{\boldmath$\nabla$\unboldmath}\times\mathcal{B}\!\!$
\mbox{\boldmath$\hat{z}$ \unboldmath}$ +\,
\mbox{\boldmath$\nabla$\unboldmath}\times\mathcal{J}\!\!$
\mbox{\boldmath$\hat{z}$ \unboldmath} and \\ $\rho_0\mathbf{v}\equiv
\mbox{\boldmath$\nabla$\unboldmath}\times\mbox{\boldmath$\nabla$\unboldmath}
\times\mathcal{W}\!\!$ \mbox{\boldmath$\hat{z}$ \unboldmath}$
+\,\mbox{\boldmath$\nabla$\unboldmath}\times\mathcal{Z}\!\!$ 
\mbox{\boldmath$\hat{z}$ \unboldmath}.
Along with the other dependent variables of the system, these potentials
are decomposed spectrally in the horizontal directions:
\begin{equation}
\mathcal{Q}=\sum_{\rm{mn}}\widetilde{\mathcal{Q}}_{\rm{mn}}(z)
   e^{2\pi i\left( f_{\rm{m}}x \,+\, g_{\rm{n}}y\right)} \, .
\end{equation}
Here, $\mathcal{Q}$ and $\widetilde{\mathcal{Q}}_{\rm{mn}}$ refer to a 
given dependent variable or scalar potential and its associated Fourier 
coefficient.  $f_{\rm{m}}$ and $g_{\rm{n}}$ are defined as 
${\rm{m}}/\mathcal{L}_x$ and ${\rm{n}}/\mathcal{L}_y$ respectively, where 
$\mathcal{L}_x$ and $\mathcal{L}_y$ denote the size of the domain in the 
$\mbox{\boldmath$\hat{x}$\unboldmath}$ and 
$\mbox{\boldmath$\hat{y}$\unboldmath}$ directions. The integers ${\rm{m}}$ 
and ${\rm{n}}$ range from $-{\rm{N}}_x/2+1$ to ${\rm{N}}_x/2$ 
and $-{\rm{N}}_y/2+1$ to ${\rm{N}}_y/2$.  ${\rm{N}}_x$ and ${\rm{N}}_y$ 
refer to the number of zones in the $\mbox{\boldmath$\hat{x}$\unboldmath}$ 
and $\mbox{\boldmath$\hat{y}$\unboldmath}$ directions.  The Fourier 
coefficients are then discretized with respect to the vertical direction 
(defined to be parallel to $\mathbf{g}$), and vertical derivatives are 
approximated by fourth-order, centered finite differences. Using
the semi-implicit method of operator splitting, we apply the 
second-order Adams-Bashforth scheme to the advection terms, and
the second-order Crank-Nicholson scheme to the diffusion terms
(see \citealt{p86} for a description of these methods). 

A detailed, in-depth description of the techniques we employ to 
obtain a solution to the anelastic equations in a non-rotating 
reference frame can be found in Appendix A of \citet{fzl99} (hereafter 
FZLF).  Only minor modifications are necessary to extend that formalism 
to apply in the case of a rotating reference frame in the modified 
\emph{f}-plane approximation. The equation for the time evolution of 
$\widetilde{\mathcal{Z}}_{\rm{mn}}$, the Fourier coefficients 
of the spectral decomposition of $\mathcal{Z}$, is obtained by 
evaluating the $\mbox{\boldmath$\hat{z}$\unboldmath}$-component 
of the curl of the momentum equation.  Thus, the addition of 
the Coriolis term to equation~(\ref{momentum}) results in an 
additional non-inertial term to equation (A16) of FZLF:
\begin{equation}
\left(\frac{\partial\widetilde{\mathcal{Z}}_{\rm{mn}}}{\partial t}
   \right)_{\rm{cor}}=\;2\Omega_z\frac{\partial\widetilde{
   \mathcal{W}}_{\rm{mn}}}{\partial z}+4\pi i \widetilde{
   \mathcal{W}}_{\rm{mn}}\left( f_{\rm{m}}\Omega_x+
   g_{\rm{n}}\Omega_y\right) \, .
\end{equation}
Here, the $\Omega_i$'s represent the Cartesian components of the angular
velocity vector, and $\widetilde{\mathcal{W}}_{\rm{mn}}$ denotes the Fourier 
coefficients of the scalar potential $\mathcal{W}$. Equation (A18) of 
FZLF results from taking the $\mbox{\boldmath$\hat{x}$\unboldmath}$ 
and $\mbox{\boldmath$\hat{y}$\unboldmath}$ components of the curl
of the momentum equation as it appears in equation (A4). Thus, we must
add another term to equation (A18) to reflect the addition of the 
Coriolis force to equation~(\ref{momentum}): 
\begin{equation}
\left(\frac{\partial\widetilde{\psi}_{\rm{mn}}}{\partial t}
   \right)_{\rm{cor}}=\;-2\Omega_z\frac{\partial}{\partial z}\left(
   \frac{\widetilde{\mathcal{Z}}_{\rm{mn}}}{\rho_0}\right) + 
   \frac{4\pi i}{\rho_0 h_0}\widetilde{\mathcal{W}}_{\rm{mn}}
   \left(f_{\rm{m}}\Omega_y-g_{\rm{n}}\Omega_x\right) - 
   \frac{4\pi i}{\rho_0}\widetilde{\mathcal{Z}}_{\rm{mn}}
   \left(f_{\rm{m}}\Omega_x+g_{\rm{n}}\Omega_y\right) \, .
\end{equation}
In the above equation, $\widetilde{\psi}_{\rm{mn}}\equiv -i(2\pi)^{-1}
(f_{\rm{m}}^2+g_{\rm{n}}^2)^{-1}(f_{\rm{m}}\widetilde{\omega}_{y\rm{mn}}
-g_{\rm{n}}\widetilde{\omega}_{x\rm{mn}})$ is the scalar variable
that corresponds to $\widetilde{\Omega}_{\rm{mn}}$ in equation
(A18) of FZLF (we change notation so as to avoid 
confusion with the components of the angular velocity
vector); $\,\widetilde{\omega}_{x\rm{mn}}$ and $\widetilde{
\omega}_{y\rm{mn}}$ are the Fourier coefficients of the 
$\mbox{\boldmath$\hat{x}$\unboldmath}$ and $\mbox{\boldmath$\hat{y}
$\unboldmath}$ components of the vorticity; and $h_0\equiv -\rho_0
(\partial\rho_0 /\partial z)^{-1}$ is the density scale height.
The last necessary modification is to equations (A23) and (A24) of
FZLF where we add the contribution from the horizontal
average of the Coriolis term in the momentum equation: 
$(\partial(\rho_0\overline{v}_x)
/\partial t)_{\rm{cor}}=2\rho_0\Omega_z\overline{v}_y$ and 
$(\partial(\rho_0\overline{v}_y)/\partial t)_{\rm{cor}}=-2\rho_0
\Omega_z\overline{v}_x$. Here, $\overline{v}_x$ and $\overline{v}_y$
denote the horizontal averages of the $\mbox{\boldmath$\hat{x}
$\unboldmath}$ and $\mbox{\boldmath$\hat{y}$\unboldmath}$ components
of the velocity (note that $\overline{v}_z$ is zero).

We now turn our attention to the non-dimensional formulation of the 
anelastic equations as implemented in the code.  As was described in 
AFF, many of the dimension-less physical variables and
fundamental physical parameters are expressed in terms of the magnetic 
field strength along the axis of the initial flux tube, $B_0$.  With 
the addition of the Coriolis force to the momentum equation, we 
introduce another fundamental dimension-less quantity, $R_b\equiv
v_{\!A}/(2\Omega H_r)\equiv 1/(2\Omega^\prime)$, which we refer to as 
the ``magnetic Rossby number'' (a quantity also discussed by
\citealt{ss92}). Here, $\Omega^\prime$ is defined as 
the non-dimensional rotation rate given in terms of the ratio of
the characteristic velocity scale to the characteristic length scale
of our problem; namely, the pressure scale height at the base
of the computational domain ($H_r$), and the Alfv\'en speed along the 
axis of the initial tube ($v_{\!A}\equiv B_0/(2\sqrt{\pi\rho_r})$, where 
the characteristic density, $\rho_r$, is taken to be the density at 
the lower boundary).  The magnetic Rossby number 
is a measure of the relative importance of the Coriolis force on 
the dynamic evolution of our buoyant magnetic flux tube.  Rotational 
effects become important when $R_b<1$, and are less significant 
when $R_b>1$.  We therefore wish to investigate three representative 
cases: $R_b=1/2$, $R_b=1$, and $R_b=2$.  Of course, the solar rotation
rate is not a free parameter --- estimates of $\Omega$ in 
different parts of the convection 
zone can be obtained from helioseismologic measurements.  Thus, 
varying the magnetic Rossby number to investigate the effect of
solar rotation on a buoyant magnetic flux tube amounts to varying 
the initial axial field strength of the tube. 

The specification of $B_0$ impacts the scaling of two of the non-dimensional
diffusive parameters: the Reynolds number ($R_e\equiv [\rho] [l] [v]
\mu^{-1}$), and the magnetic Reynolds number ($R_m\equiv [l] [v]
\eta^{-1}$). Again, the characteristic length, velocity, and density
scales ($[l]$, $[v]$, and $[\rho]$) are given in terms of 
$H_r$, $v_{\!A}$, and $\rho_r$ respectively.  For this study, 
we hold the coefficients of viscosity and 
magnetic diffusion ($\mu$ and $\eta$ respectively) constant between runs 
(where possible), and vary $B_0$ such that we obtain our three representative 
values of the magnetic Rossby number. In this manner, we focus our attention 
on the effects of solar rotation on the dynamic evolution of a rising 
$\Omega$-loop.  We note however, that as $B_0$ becomes large, it is
increasingly difficult to hold the coefficient of viscosity constant
for a given grid resolution.  This is because as the Reynolds number
increases, features develop at the forefront of the rising tube and in the
trailing eddies that are on the order of a few grid zones in size.
If this occurs, the code becomes unstable, and the grid resolution must 
be increased to alleviate the problem.  Table~\ref{tbl-1} lists the 
values of the relevant physical parameters for each of seven simulations.
For the cases where $B_0\sim 10^5$G, values of $\mu$ and $\eta$ were
increased to avoid instabilities. We otherwise would have to
double the grid resolution in the tube cross-section to take full 
advantage of the FFT algorithms employed in the code, and such a large
increase is unnecessary for the purposes of this study.  

Our labeling convention (see column one of Table~\ref{tbl-1}) consists 
of four letters which signify both the strength of $B_0$, and the latitude 
of the computational box. For example, LFHL refers to the run which 
has an initial magnetic flux tube with a relatively \emph{L}ow axial 
\emph{F}ield strength positioned at a \emph{H}igh \emph{L}atitude, while
HFLL refers to the run with an initial tube of relatively \emph{H}igh 
axial \emph{F}ield strength positioned at \emph{L}ow \emph{L}atitude 
--- and so on.  Run SS0 refers to the simulation where solar rotation
was neglected (this notation was chosen in order to be consistent
with the naming convention of AFF).  In each case, we begin with a static,
horizontal, cylindrical magnetic flux tube positioned near
the bottom of a simulation box that spans 5.147 pressure scale 
heights vertically.  The initial field configuration is given by
$\mathbf{B}=B_x(r)\mbox{\boldmath$\hat{x}$\unboldmath}$, where 
$B_x(r)\equiv B_0e^{-r^2/a^2}$ and $r$ is the radial distance to 
the central axis in the tube's cross-section.  The size of the initial
tube, $a$, is taken to be $0.1H_r$ in each of the simulations.  We 
choose not to include any initial field line twist about the central 
axis of the tube, since we wish to determine what role the Coriolis 
force \emph{alone} plays in the overall fragmentation of the tube 
as it rises toward the photosphere.  As was done in AFF, we apply an 
artificial entropy perturbation to the midsection of the tube at time 
$t=0$. This causes the tube to rise as an $\Omega$-loop without having 
to take the computationally expensive step of evolving each case 
self-consistently from a state of initial force balance (as was done
in \citealt{cms95,ff96,f00}). 

\section{Results}\label{results} 

Each simulation begins with an untwisted cylindrical magnetic
flux tube embedded near the base of our model convection zone.
An entropy perturbation is applied to the flux tube such that in
the absence of rotation, the tube will rise as an $\Omega$-loop,
and the apex will approach the photospheric boundary in a highly
``fragmented'' state.  That is, a majority of the magnetic flux
will be concentrated away from the tube's central axis along
vortex pairs that are shed during the tube's ascent.
Following AFF, we consider a section of the $\Omega$-loop to be 
fragmented if the ratio of the magnetic field weighted second 
moments of position along the Frenet binormal and normal directions 
of the tube exceed $1.5$ (see Section 3 of AFF for details on
how tube fragmentation is determined, and see Figures 1 and 3 in 
AFF for visual examples of a fragmented and non-fragmented 
$\Omega$-loop).  We expect that solar rotation will
play a dynamically important role during the tube's ascent, since 
the loop's rise time is on the order of a rotation period 
for reasonable initial field strengths ($3\times 10^4$G $\lesssim
B_0 \lesssim 1\times 10^5$G, see \citealt{ffl00}).  

\begin{figure}
\centerline{\epsfig{file=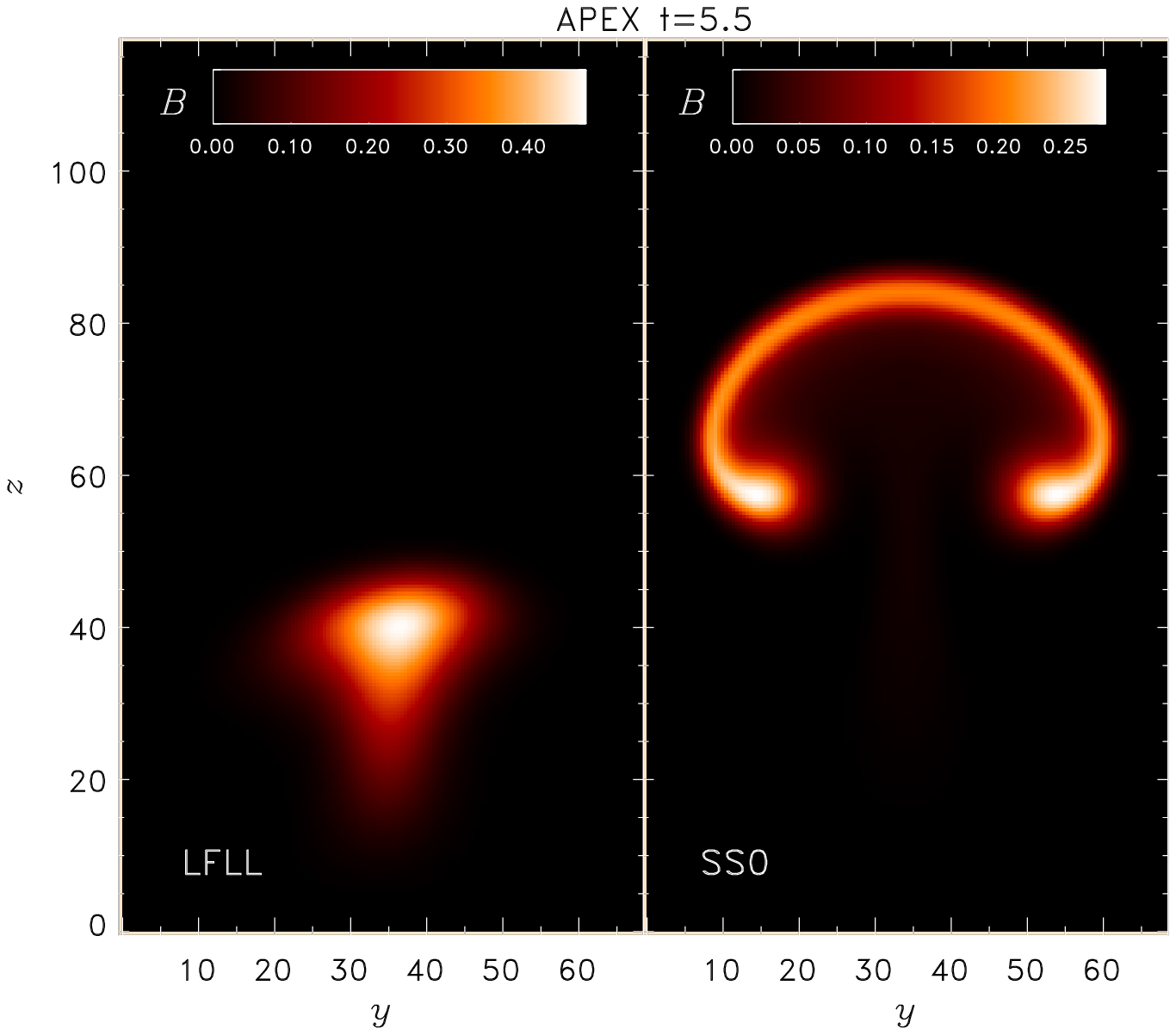}}
\caption{Apex cross-sections of $|\mathbf{B}|$ for a
simulation with $R_b=0.5$ at $15^\circ$ latitude (left frame,
run LFLL), and a simulation neglecting solar rotation (right
frame; labeled run SS0 to be consistent with the naming
convention of AFF). $|\mathbf{B}|$ is given in terms of the field
strength
along the axis of the initial tube, and $t$ is given in units of
the ratio of the pressure scale height at the base of the
domain to the Alfv\'en speed along the axis of the initial tube.
The coordinate axes are given in terms of computational zones,
though only a portion of the domain is shown. \label{fig1}}
\end{figure}

\begin{figure}
\centerline{\epsfig{file=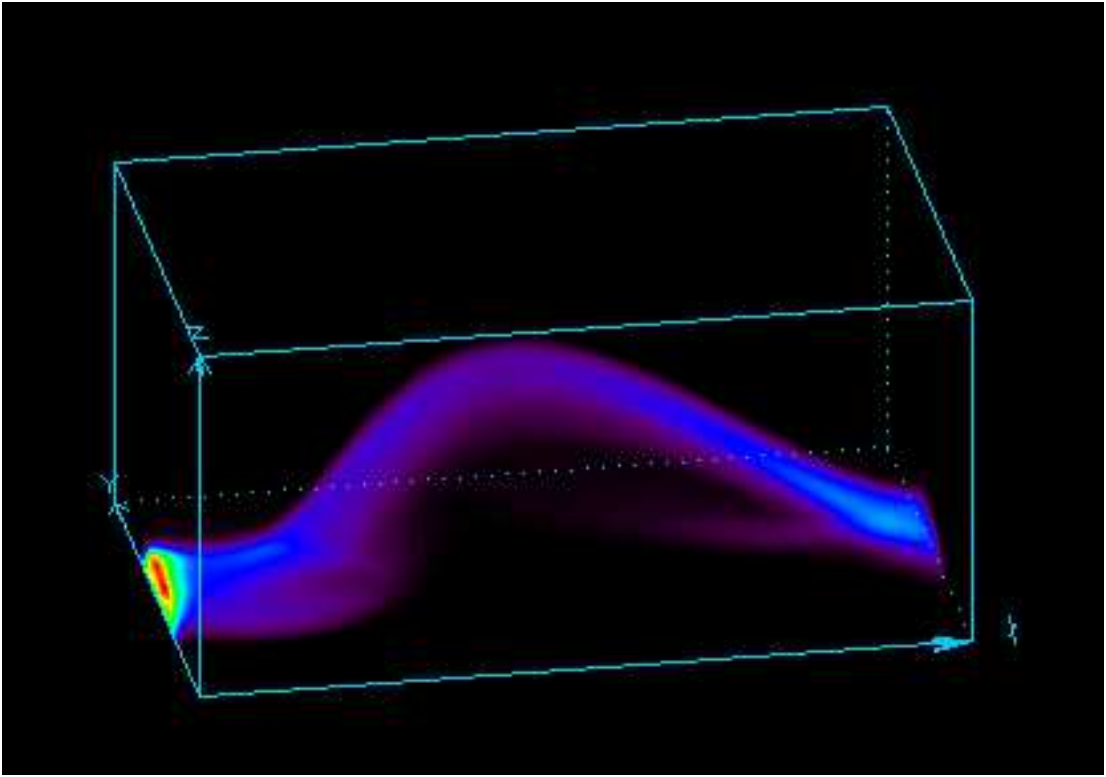}}
\caption{A volume rendering of $|\mathbf{B}|$ for
the $\Omega$-loop of run HFLL at $t=8.53$.  The entire rectangular
computational domain is shown. \label{fig2}}
\end{figure}

We find that the impact of the Coriolis force on a rising flux tube is 
dramatic, as demonstrated in Figure~\ref{fig1}.  In this figure, 
we compare the magnetic field strength in the apex cross-section 
of two $\Omega$-loops:  one which rises at $15^\circ$ latitude in the 
presence of solar rotation ($\Omega \sim 13.9^\circ$/day) with
an initial axial field strength of $B_0=2.67\times 10^4\,$G 
(corresponding to $R_b=0.5$), and an identical loop which rises in the 
absence of rotation.  Two effects are immediately apparent.  First, 
the $\Omega$-loop that rises through the rotating atmosphere has yet to
attain the height of the loop that rises through the non-rotating 
atmosphere; and second, the loop subjected to the effects of the
Coriolis force retains its cohesion, and does not fragment. 
Figure~\ref{fig2} is a volume rendering of a rising magnetic flux 
tube in a run where $R_b=2$ and $B_0=1.0\times 10^5\,$G.  
The coordinate axes are chosen 
such that \boldmath $\hat{y}$ \unboldmath points northward, 
and \boldmath $\hat{z}$ \unboldmath points radially outward.  
Again, we see that the tube retains its cohesion during its ascent.  
Figure~\ref{fig2} also demonstrates that in the presence of rotation, 
the $\Omega$-loop is somewhat asymmetric about the apex. That is, the 
trailing portion of the loop forms a steeper angle with respect to
the \boldmath $\hat{x}$ \unboldmath direction than the leading
side of the loop.

\begin{figure}
\centerline{\epsfig{file=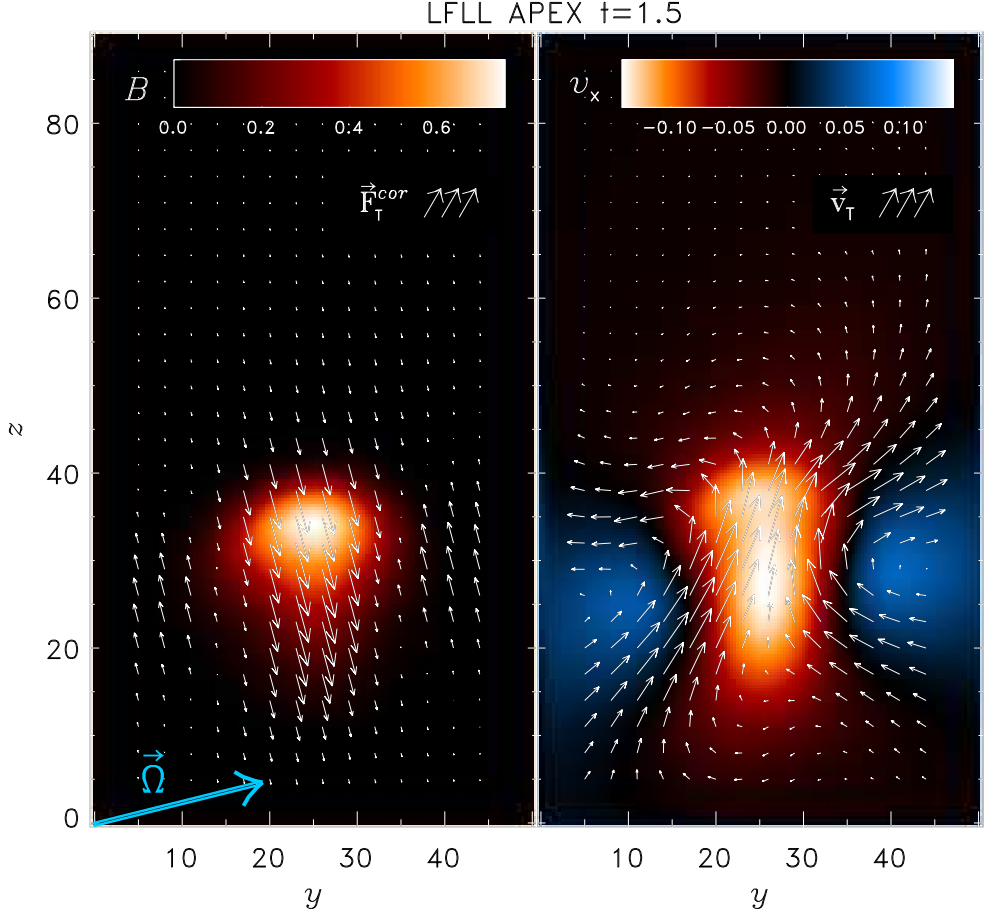}}
\caption{Apex cross-sections of $|\mathbf{B}|$
along with the transverse Coriolis force (left frame), and the
corresponding velocity field (right frame) for run LFLL
($R_b=0.5$ at $15^\circ$ latitude) at $t=1.5$. Quantities are
given in naturalized units (see text for details).
In the right frame, axial flows depicted in red indicate
velocities receding from the viewer, while blue
colors indicate flow moving toward the viewer.
Here, max$|\mathbf{v}_x|/$max$|\mathbf{v}_T|\approx 1.5$;
that is, the axial flows are stronger than the transverse
flows.  The angular velocity vector is displayed in the lower
left-hand corner of the left frame, and applies to the entire
figure. \label{fig3}}
\end{figure}

A qualitative understanding of these effects can be obtained
by examining Figure~\ref{fig3}, which shows $|\mathbf{B}|$, 
the transverse components of the Coriolis force, and the 
flow field in the apex cross-section during the initial stages
of run LFLL (an earlier snapshot of the same run that is displayed 
in the left frame of Figure~\ref{fig1}).  In this plot 
(and all others, unless otherwise stated) the physical 
quantities are displayed in naturalized units: the 
magnetic field strength is given in terms of the strength 
of the field along the axis of tube at $t=0$ ($B_0$), 
the velocities are given in terms of the Alfv\'en speed along
the axis of the initial tube ($v_{\!A}$), and the unit of time is
the ratio of the pressure scale height at the base of the 
computational domain to the characteristic velocity, $H_r/v_{\!A}$.  
For run LFLL, shown in Figure~\ref{fig3}, $B_0=2.67\times 10^4\,$G,
$v_{\!A}=1.68\times 10^4$ cm s$\!\,^{-1}$ (for 
$\rho_r=0.2$ g cm$\!\,^{-3}$), and $t=1.5$ corresponds to 
$6.18$ days.  The axes in the figure are given in terms of
computational zones; in this case, each individual zone is 
$0.017 H_r$, or approximately $1$ Mm.  

In the right-hand frame of Figure~\ref{fig3} we see that a 
strong, retrograde axial flow has developed as a result 
of the Coriolis force acting on rising material within the buoyant 
loop.  This is in sharp contrast to the simulations in AFF, where 
solar rotation is neglected and the initial flux tube has little 
or no field line twist.  In these cases, the strong axial flow 
is absent, and counter-rotating vortex pairs can quickly form on 
either side of the tube's central axis (see also 
\citealt{wmhp00,dn98}).  As was shown in AFF and \citet{lfa96}, 
the non-vertical component of the hydrodynamic ``lift'' force 
per unit volume generated by flows in the vortex tubes acts to push 
them apart, ultimately resulting in fragmentation near the apex 
of the $\Omega$-loop.  However, in the presence of rotation, we 
find that the circulation (and thus the degree of fragmentation 
of the loop) is substantially reduced.  Roughly speaking, 
this follows from the fact that the Coriolis force can do
no work (it always acts in a direction perpendicular to the 
flow); it simply redistributes kinetic energy between transverse 
and axial flows.  The initial increase in the axial flow at the 
apex of the loop gives rise to a secondary effect: a transverse 
component of the Coriolis force normal to the angular velocity 
vector.  At low latitudes, this component acts primarily in the 
$-$\boldmath $\hat{z}$ \unboldmath direction, and opposes the 
buoyant rise of the loop; however, it also forces the top of the 
loop to drift northward over time. 

In simulations where the effects of rotation are neglected,
we find that material flows along the rising tube away from the 
apex.  In the presence of rotation, this type of ``draining'' flow 
manifests itself as an asymmetry in the axial flow induced by the 
Coriolis force: the retrograde flow is stronger in the trailing
side of the loop than it is in the leading side.  As the Coriolis 
force acts upon these asymmetric axial flows, and as the buoyant,
magnetized plasma attempts to preserve its angular momentum, the 
morphological asymmetry predicted by \citet{mcs94} and apparent in 
Figure~\ref{fig2} quickly develops.

\begin{figure}
\centerline{\epsfig{file=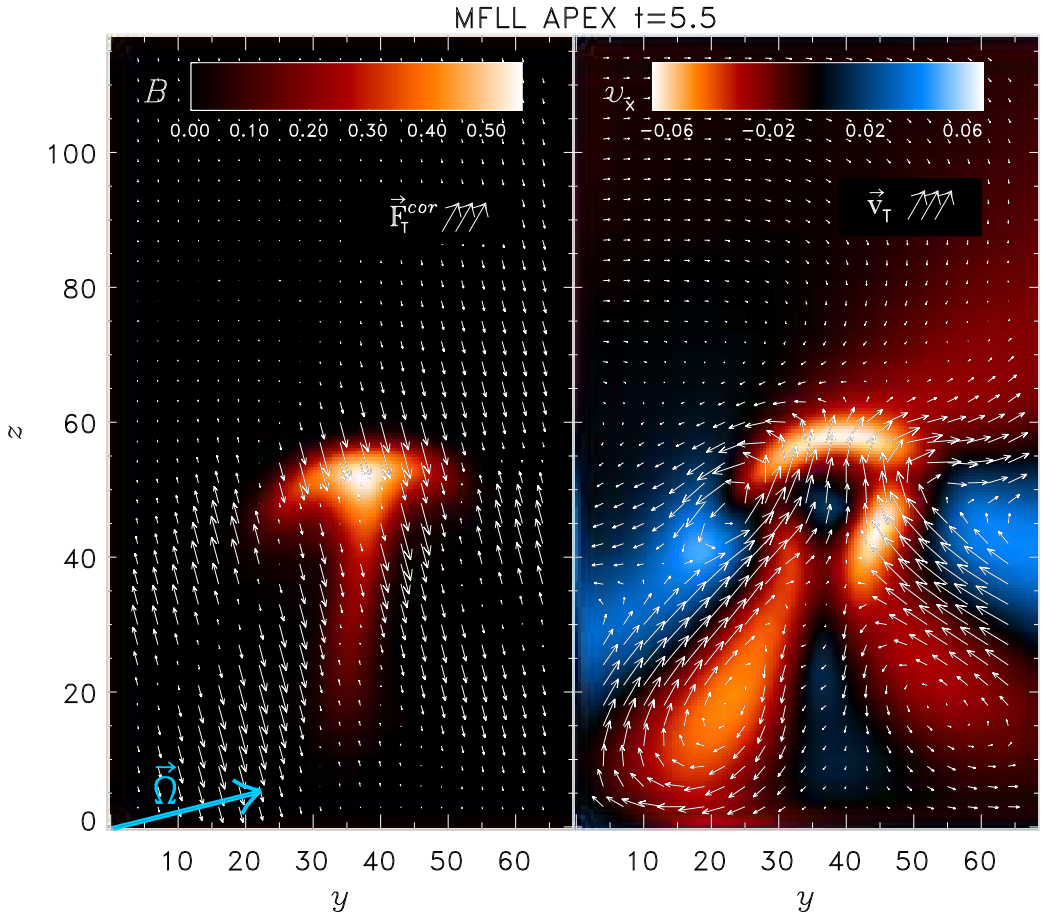}}
\caption{Same as Figure~\ref{fig3} for run
MFLL ($R_b=1$ at $15^\circ$ latitude) at $t=5.5$
(the same time as displayed in Figure~\ref{fig1}).
Here, max$|\mathbf{v}_x|/$max$|\mathbf{v}_T|\approx 1.1$.
\label{fig4}}
\end{figure}

Figure~\ref{fig4} shows the transverse components of the 
Coriolis force, and the flow field in the apex cross-section
of run MFLL at $t=5.5$ (the same time, in naturalized 
units, as the cross-sections of Figure~\ref{fig1}). In this 
run, $R_b=1$, and the domain is again centered at $15^\circ$ latitude.   
This implies that $B_0=5.35\times 10^4\,$G, $v_{\!A}=3.37\times
10^4\,$cm s$\!\,^{-1}$, and the unit of time is 2.06 days ($t=5.5$
thus corresponds to $11.3$ days).  In an earlier snapshot of 
run MFLL ($t=1.5$), the morphology and flow patterns are very 
similar to those shown for run LFLL in Figure~\ref{fig3}.  However,
as the runs progress, differences in secondary and tertiary
flows become more pronounced, and the loops begin to evolve
in a dissimilar manner. This is evident when comparing the 
first frame of Figure~\ref{fig4} with the frames of 
Figure~\ref{fig1}, which highlight the striking differences in 
morphology that ultimately develop between simulations with 
different values of the magnetic Rossby number.  These figures
show that as the initial axial magnetic field strength of the 
tube is increased, the tendency for the tube to fragment as
it rises toward the photosphere is also increased.

\subsection{An Approximate Model}

To better understand the relationship between magnetic 
Rossby number, latitude of flux emergence, and loop fragmentation,
we approximate the forcing terms of the momentum equation
[equation~(\ref{momentum})] by constant, order-of-magnitude
estimates evaluated for our initial conditions, and valid
for the initial stages of evolution.  We then
have a simplified expression for the velocity field that
admits to an analytic solution:
\begin{equation}
\frac{d\mathbf{v}}{dt}=-2\left(
   \mbox{\boldmath$\Omega$\unboldmath}
   \times\mathbf{v}\right) + \mathbf{Q} \, . 
\end{equation}
It is important to recognize that this simple approximation
will not capture the factor of two reduction in acceleration
that occurs in the non-rotating case because of ``added inertia''
effects (see \citealt{lfa96}).  Nevertheless, it does capture
the essence of the dynamic effects introduced by the Coriolis force.
Let $(q_x,q_y,q_z)$ represent the Cartesian components of the
constant driving term $\mathbf{Q}$, and let $\theta$ represent
solar latitude.  Then the general solution to the above equation 
is given by
\begin{eqnarray}
   v_x(t) &=& \frac{q_x}{2\Omega}\sin{2\Omega t} -
   v_x^{(P)}\cos{2\Omega t} + v_x^{(P)} \\
   v_y(t) &=& \frac{q_x}{2\Omega}\sin{\theta}\cos{2\Omega t} +
     v_x^{(P)}\sin{\theta}\sin{2\Omega t} \nonumber \\
     &\;& \;\;\;\;\;\;\;\;+ \;(q_y-2\Omega v_x^{(P)}\sin{\theta})t -
     \frac{q_x}{2\Omega}\sin{\theta} \\
   v_z(t) &=& -\frac{q_x}{2\Omega}\cos{\theta}\cos{2\Omega t} -
     v_x^{(P)}\cos{\theta}\sin{2\Omega t}  \nonumber \\
     &\;& \;\;\;\;\;\;\;\;+ \;(q_z+2\Omega v_x^{(P)}\cos{\theta})t +
     \frac{q_x}{2\Omega}\cos{\theta} \, ,
\end{eqnarray}
where
\begin{eqnarray}
   v_x^{(P)}&=&\frac{1}{2\Omega}(q_y\sin{\theta}-q_z\cos{\theta})
     \, . 
\end{eqnarray}

\begin{figure}
\centerline{\epsfig{file=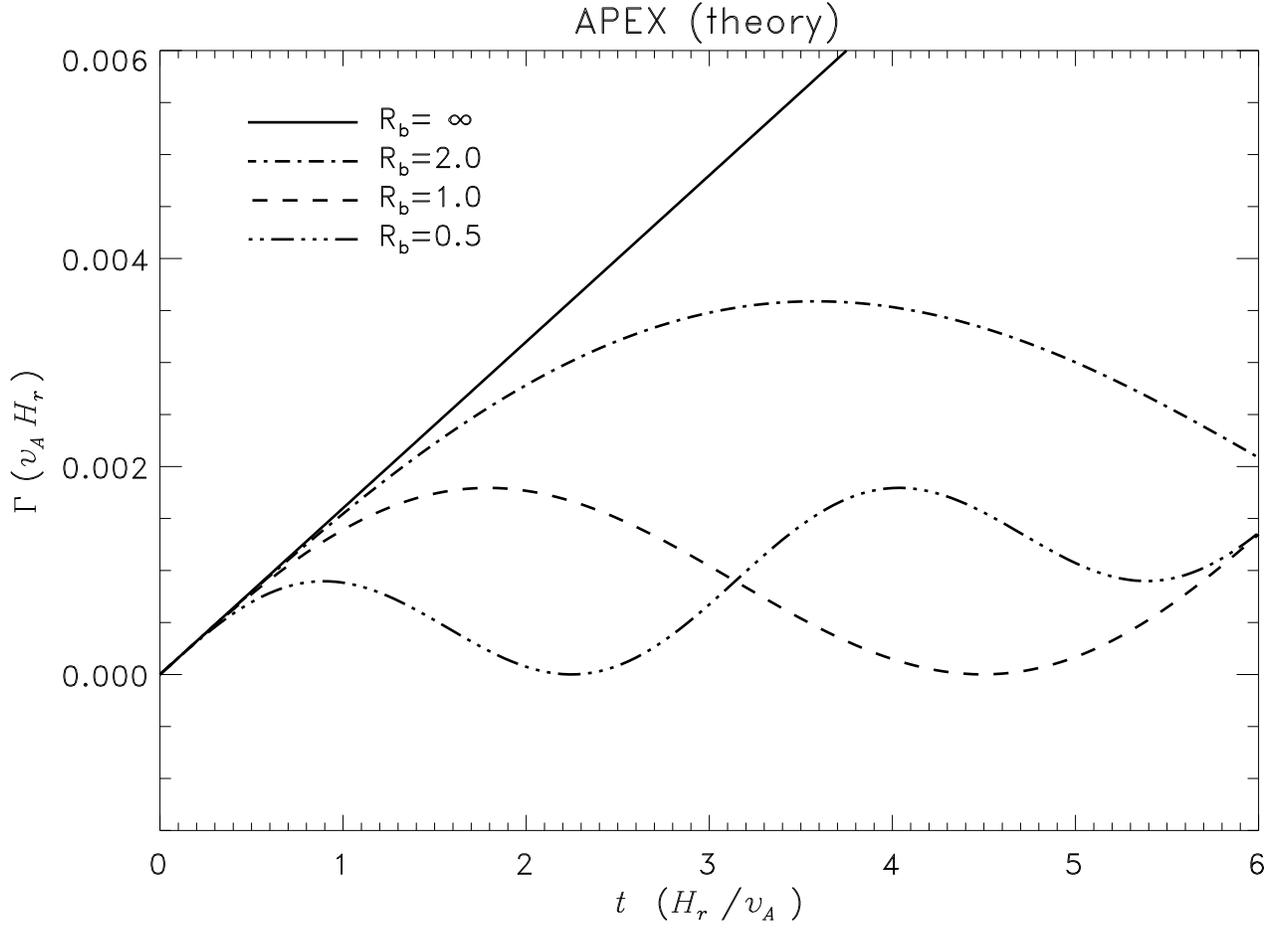}}
\caption{Analytic solution for the total
circulation of a tube ``fragment'' as a function of
time for four representative values of the magnetic
Rossby number.  The quantities are given in naturalized
units.  \label{fig5}}
\end{figure}

The total circulation over a cross-sectional slice of the flux
tube is given by $\Gamma=
\int_S\,$\boldmath$\omega$\unboldmath$\,\cdot\,\hat{\mathbf{n}}_S\,
dS= \oint_C\mathbf{v}\cdot d$\boldmath$\ell$\unboldmath$\,$,
where $C$ is the closed circuit bounding the cross-sectional surface
$S$, $\hat{\mathbf{n}}_S$ is the outward normal to $S$,
and \boldmath$\omega$\unboldmath$=$\boldmath$\nabla$\unboldmath
$\times\mathbf{v}$ is the vorticity vector.  
As the tube begins to rise, oppositely directed
circulation patterns form on either side of a line defined by the
Frenet normal.  The total circulation of these regions (at the loop
apex) is given by $\Gamma=\oint_C \mathbf{v}\,\cdot
d$\boldmath$\ell$\unboldmath$\,\,\,\approx\,\int_{\ell_a}
d\ell_a \, v_z\, \approx\, a\!<\!v_z\!>$, if we assume
that the primary contribution to $\Gamma$ stems from
$\mathbf{v}\cdot d$\boldmath$\ell$\unboldmath$\,$ across the
diameter of the tube, $a$.  Since $v_z$ is roughly constant across
the interior of the tube, we can approximate the circulation (and
thus the forces acting to pull the tube apart) with our analytic
solution for the velocity.  Initially, the motion of the tube
is determined by the buoyancy force, which acts in the
\boldmath $\hat{z}$ \unboldmath direction.  Thus, we assume that
$q_x$ and $q_y$ are zero, and express the circulation as:
\begin{equation}
   \Gamma(t)=\frac{{a v_{\!A}}^{\!2}}{2H_r}
   \left(t\sin^2\theta+\frac{\sin(2\Omega t)}{2\Omega}
   \cos^2\theta\right) \, . 
\end{equation}
We plot our simple approximation of $\Gamma(t)$ for four
representative values of $R_b$ in Figure~\ref{fig5}.  

\begin{figure}
\centerline{\epsfig{file=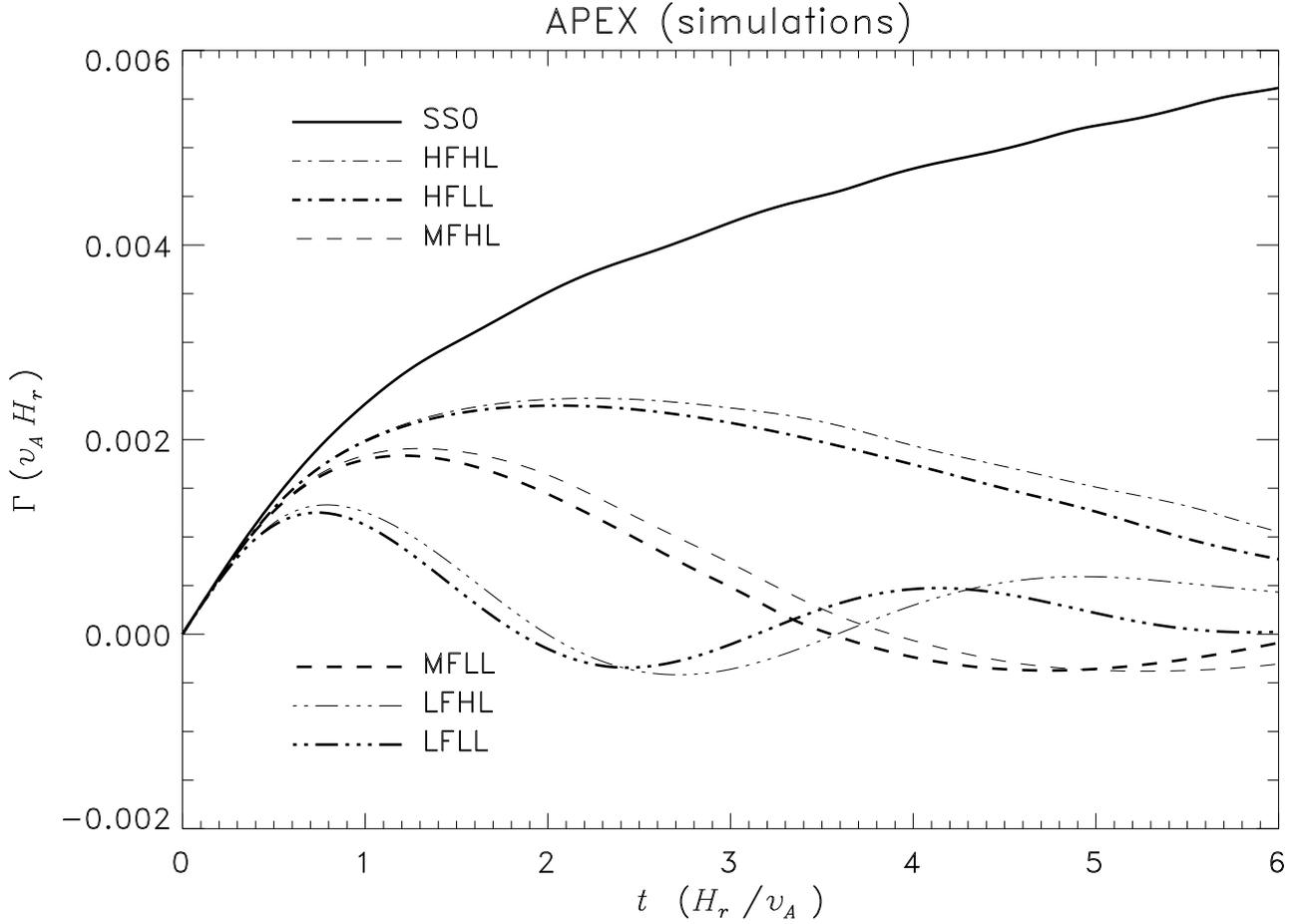}}
\caption{3-D MHD solution for the total circulation
of a tube ``fragment'' at latitudes of $15^\circ$ and $30^\circ$
as a function of time and magnetic Rossby number.  The quantities
are given in naturalized units.
\label{fig6}}
\end{figure}

In order to compare our approximation with the 
MHD solution, we must first obtain $\Gamma(t)$ directly from the 
simulation data.  To accomplish this, 
we follow the methodology described in Section 3 of AFF, 
and first calculate the path of the magnetic flux tube through 
the background plasma.  The position of the loop (and its 
fragments, if applicable) is specified by the 
magnetic field weighted first moments of position along a series 
of two-dimensional slices that closely 
correspond to the tube's cross-sectional plane.   Once the 
path of the tube is determined (see Section 3 of AFF), 
we construct the Frenet basis vectors, 
and define two regions on either side of a line defined by 
the Frenet normal where the oppositely directed circulation patterns 
will develop.  From the simulation results, 
we then compute the total circulation 
across each of these regions.  Figure~\ref{fig6} 
shows $\Gamma(t)$ for each simulation.  One can
compare these results with the simple approximation shown in 
Figure~\ref{fig5}.  It is clear that the general pattern
of time dependence of the circulation is successfully reproduced
by our simple model, though there are significant differences in
the details.

The oscillations in $\Gamma(t)$ are inertia waves (see \citealt{b67})
with a period of $t=2\pi R_b$ (here, $t$ is expressed in
normalized units of $H_r/v_{\!A}$). Thus, the tendency for the loops
to fragment is significantly reduced after the circulation begins
to fall off at $t=\pi R_b/2$.  The overall evolution of the
loop depends strongly on the dynamics during the initial portion of
its rise, since the early differences in secondary and tertiary
flows induced by the Coriolis force have a significant impact
on the eventual evolution of the tube.  We find that if the
tube is able to rise several of its diameters before $t$
reaches $\pi R_b/2$, then the resulting $\Omega$-loop will
ascend toward the photosphere in a fragmented state ---
otherwise, the tube evolves cohesively.  

We can infer from our approximate model that, on average, the 
rise speed of the tube slowly increases over time.  This prediction
is confirmed by the simulations, which (for sufficiently high
choices of $R_e$ and $R_m$) show the initial slow, linear 
increase in the rise speed of the tube, modulated by inertial 
oscillations.  This behavior persists in the simulations until
$t$ becomes comparable to the diffusive timescale.  If this occurs, 
magnetic diffusion begins to (artificially) reduce the buoyancy 
of the tube, eventually preventing it from 
approaching the upper boundary (this occurred in runs LFLL and
LFHL after $t\approx 34.0$). Of course, this problem can be
mitigated by reducing the coefficient of magnetic diffusivity
in the simulations, but this requires a significant increase in 
grid resolution (to avoid numerical instabilities); and since our 
analysis concentrates on the early stages of loop evolution, such 
a computationally expensive step is unwarranted.

\subsection{Comparison with Two-dimensional Models}

\begin{figure}
\centerline{\epsfig{file=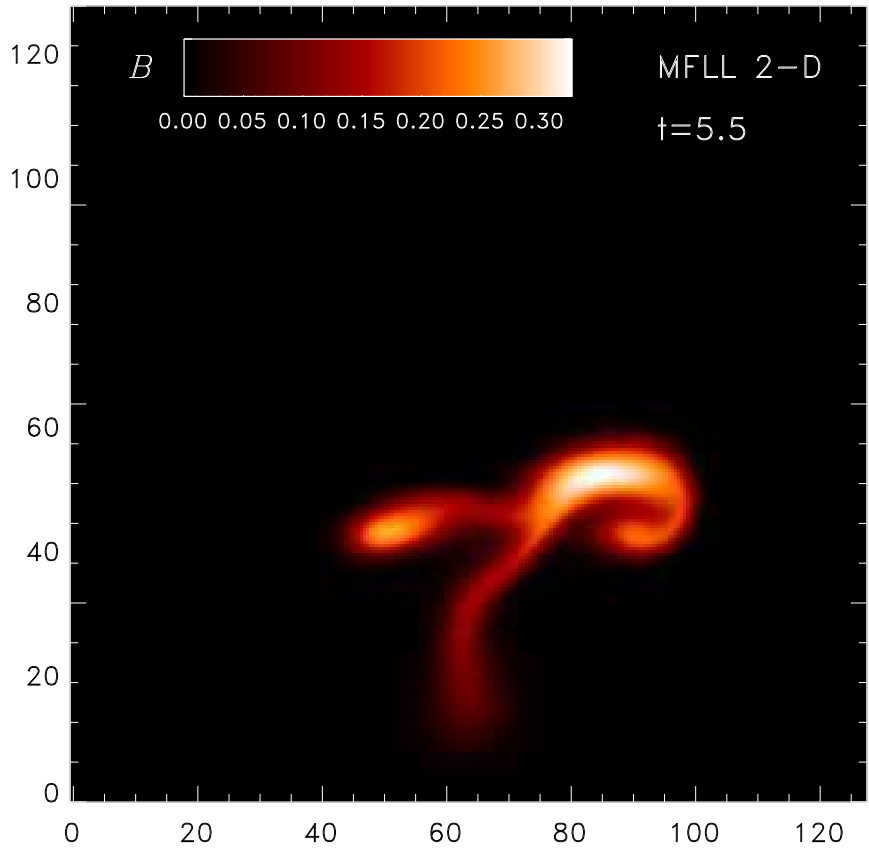}}
\caption{A cross-section of $|\mathbf{B}|$
at $t=5.5$ (in normalized units) for a 2-D run that
begins with the initial magnetic flux tube of run MFLL.
\label{fig7}}
\end{figure}

To investigate the role that three-dimensional geometry plays 
in the tube's evolution, we compare our runs with simulations
of buoyant, axially symmetric magnetic flux tubes moving
in two dimensions. Figure~\ref{fig7} shows the tube 
cross-section of a 2-D run whose parameters are that of 
the 3-D run MFLL shown in the first frame of 
Figure~\ref{fig4}.  This 2-D case can be thought of as an 
infinitely-long, buoyant, axially symmetric $\Omega$-loop of 
zero apex curvature.  A comparison of the apex cross-section
shown in Figure~\ref{fig4} with the cross-section of
Figure~\ref{fig7} reveals that a 2-D geometric constraint
imposed on the solution (even in the early stages
of loop evolution) significantly alters the dynamic
evolution of the tube.  This suggests that in the
presence of solar rotation, the geometry of the $\Omega$-loop 
still plays an important role in determining the morphology of
the emerging magnetic field --- a result consistent with the 
findings of AFF for similar runs where solar rotation was 
neglected.  

In 2-D, the Coriolis force still acts to
suppress the fragmentation of the tube. However, the
hydrodynamic ``lift'' force that acts to fragment the tube
is overestimated due to the tube's effectively 
infinite axial length scale. It is
the binormal component of this force that acts to pull the tube 
apart (see \citealt{lfa96,fzl98}; AFF), thus the overall tendency
for the tube to retain its cohesion during its ascent is
reduced as compared to the 3-D cases. In addition, 
the amount that the tube is deflected toward
the pole is significantly overestimated in a 2-D geometry.  This
is because in 3-D, the component of the Coriolis force that
acts to push the tube poleward acts only on the portion 
of the $\Omega$-loop where there are strong, mainly axial
flows (eg. near the apex).  In general, our calculations support 
the conclusions of \citet{wmhp00} who used 2-D simulations to
demonstrate that in the presence of the Coriolis force,
flux tubes remained somewhat more cohesive during their ascent.
We note, however, that in 3-D, this effect is more dramatic.

\subsection{Morphology of the Emerging Magnetic Field}\label{morphology}

There are several reasons why we must be cautious when making 
predictions about the behavior of emerging magnetic flux based 
on data obtained from our simulations.  First, the anelastic
approximation becomes marginal as the flux tube approaches the
photospheric boundary, where densities are lower, and the
acoustic Mach number of the flow approaches unity.  Second,
we intended for these simulations to investigate the effects
of the Coriolis force alone on the evolution of our 
$\Omega$-loop.  Thus, we did not include any field
line twist along our initial tube, nor did we consider the effects 
of a tube rising through a dynamically convecting background
state. Third, our models do not account for radiative losses 
that occur at the surface.  However, given these caveats,
we feel that we can still make some general predictions 
regarding expected magnetic field and velocity flow patterns 
if the magnetic flux emerges through a relatively uncluttered 
region on the solar disk.

\begin{figure}
\plotone{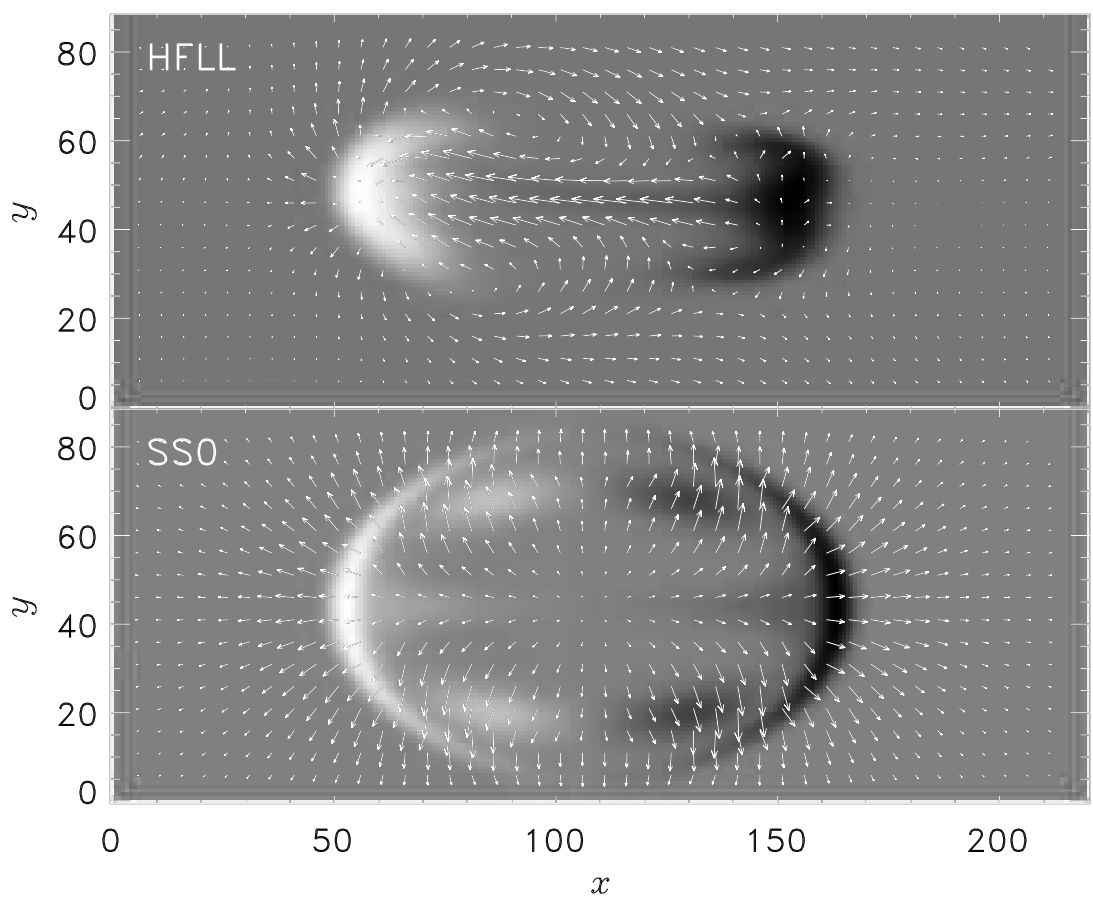}
\caption{A grey-scale image of the radial magnetic
field for a horizontal slice just below the apex of the
$\Omega$-loops of the final timesteps of runs HFLL
(top frame) and SS0 (bottom frame).  The vector field
represents the transverse ($x$-$y$) flow field.  The
$x$ and $y$ scales are given in terms of computational
zones.  The grid is uniform; each zone is roughly $1$ Mm
in size. There is a very slight tilt of approximately
$2.5$ degrees from horizontal between the center of the
bipolar regions.
\label{fig8}}
\end{figure}

Figure~\ref{fig8} shows the radial component
of the magnetic field along horizontal slices taken
underneath the apex of the rising $\Omega$-loops of runs
HFLL, and SS0.  This figure emphasizes the differences in 
morphology of emerging magnetic field between a simulation 
that includes the effects of solar rotation (run HFLL),  
and one that does not (run SS0).  In the figures,
outwardly directed field is displayed in white, and inwardly 
directed field is displayed in black.  Superimposed over the 
each grey-scale image is the transverse velocity field.  
One of the most striking effects of the Coriolis force
is the strong flow that rapidly develops in both the magnetized and
unmagnetized plasma directed from the leading toward
the trailing polarity (the leading polarity moves in the 
direction of solar rotation as the loop continues to rise).  
This flow is simply the result of the Coriolis force acting on 
the rising fluid just below the apex of each loop.  This
``LTT'' (\emph{L}eading \emph{T}o \emph{T}railing polarity) 
flow is evident in all of our simulations except in cases 
where solar rotation is neglected.

The Coriolis force manifests itself in several other 
interesting ways.  First, since the trailing edge of the loop
emerges at a steeper angle with respect to the surface than 
the leading edge (an effect most pronounced in the runs where 
$R_b$ is small), the leading polarity becomes 
``spread-out'' in the east-west direction (along $x$).  
Second, in the cases where $R_b$ is small, we observe that the
leading polarity is positioned slightly closer to the equator
than the trailing polarity --- behavior that is roughly 
consistent with the ``Joy's Law'' result from thin flux tube models 
(see \citealt{dc93,ffm94,s94,cms95,ff96,ffl00}). 
Third, the north-south dispersion of the emerging flux is greatly reduced
when solar rotation is included in the simulations.  
This reflects the increased cohesion of the flux rope during its
ascent through our model convection zone.

\section{Summary and Conclusions}\label{conclusions}
AFF found that the degree of fragmentation of a rising
$\Omega$-loop depends upon the 3-D geometry of the loop --- the
greater the apex curvature, the lesser the degree of fragmentation
for a fixed amount of initial field-line twist.  We are now able
to extend this analysis to a rotating model convection zone, and
investigate how the degree of apex fragmentation depends on the
axial field strength of the initial magnetic flux tube.  Assuming
a given rotation rate, we find that: 
\begin{enumerate}
\item A magnetic flux tube that rises buoyantly toward the
photosphere in a rotating convection zone rises more slowly, 
and is better able to retain
its cohesion than an identical flux tube that rises through a
non-rotating convection zone --- even in the absence of initial
field line twist.  If the tube is sufficiently buoyant so that it 
is able to rise a distance of several tube diameters before 
$t=\pi/4\Omega$ (where $\Omega$ is an estimate of the relevant 
solar rotation rate at a given latitude), then the tube 
will fragment during its rise toward the photosphere.
\item In the absence of forces due to convective motions, 
the stronger the initial axial field strength of the tube,
the more fragmentation occurs as the tube rises.
\item An approximate model is able to predict the
initial evolution of the circulation, and hence can effectively
characterize the formation and subsequent interaction of the
oppositely-directed vortex pairs of a fragmented $\Omega$-loop.
\item One must be careful not to over-interpret results 
from two-dimensional simulations of buoyant magnetic flux tubes.
In a two-dimensional geometry, hydrodynamic forces due to vortex interaction
are overestimated.  The amount of poleward deflection 
due to the Coriolis force is also overestimated. 
\item If magnetic flux emerges as an $\Omega$-loop, the Coriolis force 
will induce a relatively strong flow that is directed from the
region of leading polarity to the region of trailing polarity.
This flow is present in both the magnetized and unmagnetized plasma. 

\end{enumerate}

\clearpage

\acknowledgments
This work was funded by NSF grants AST 98-19727 and ATM 98-96316, by
NASA grant NAG5-8468, and by ONR grant FDN00173-00-1-G901-03/03.  
The computations described here were supported by
the National Center for Atmospheric Research under grant ATM 98-96316.
Further computational support was provided by the National Computational
Science Alliance. We would like to thank the authors of the FFTW package 
\citep{fj97} for making their code publicly available.

\clearpage

\clearpage

\clearpage

\clearpage

\clearpage

\clearpage

\clearpage

\clearpage

\clearpage
 

\begin{deluxetable}{cccccccccccc}
\footnotesize
\tablecaption{Physical Parameters \label{tbl-1}}
\tablewidth{0pt}
\tablehead{
\colhead{Label} & \colhead{Resolution\tablenotemark{a}}
   & \colhead{Lat\tablenotemark{b} $\:$($\Omega$)\tablenotemark{c}} 
   & \colhead{$B_0\,(G)$} & \colhead{$R_b$} & \colhead{$R_e$,$R_m$} 
   & \colhead{$\mu$,$\nu$ (cgs)} 
}
\startdata
LFHL & 256$\times$128$\times$128 & 30 (13.5135)  & 2.5966$\times 10^{4}$ 
   & 0.5 & 1700.07 & 5.7805$\times 10^{10}$ \\
LFLL & 256$\times$128$\times$128 & 15 (13.9104)  & 2.6728$\times 10^{4}$ 
   & 0.5 & 1750.00 & 5.7805$\times 10^{10}$ \\
MFHL & 256$\times$128$\times$128 & 30 (13.5135)  & 5.1932$\times 10^{4}$ 
   & 1.0 & 3400.15 & 5.7805$\times 10^{10}$ \\
MFLL & 256$\times$128$\times$128 & 15 (13.9104)  & 5.3457$\times 10^{4}$ 
   & 1.0 & 3500.00 & 5.7805$\times 10^{10}$ \\
HFHL & 256$\times$128$\times$128 & 30 (13.5135)  & 1.0386$\times 10^{5}$ 
   & 2.0 & 3400.15 & 1.1561$\times 10^{11}$ \\
HFLL & 256$\times$128$\times$128 & 15 (13.9104)  & 1.0691$\times 10^{5}$ 
   & 2.0 & 3500.00 & 1.1561$\times 10^{11}$ \\
\enddata
\tablenotetext{a}{($x$,$y$,$z$)}
\tablenotetext{b}{Latitude, degrees}
\tablenotetext{c}{Observed rotation rate, degrees/day --- data from 
   MDI 2dRLS inversion \citep{s97}}
\end{deluxetable}

\end{document}